\documentclass[graybox]{svmult}

\usepackage{makeidx}         
\usepackage{graphicx}        
\usepackage{multicol}        

\makeindex             

\begin{document}

\title*{Structure formation in the local Universe and the cosmological constant}
\author{V.G. Gurzadyan}
\institute{Center for Cosmology and Astrophysics, Alikhanian National Laboratory and Yerevan State University, \at Yerevan 0036, Armenia, \email{gurzadyan@yerphi.am}
}
%
%
\maketitle

\abstract*{The structure formation in the local Universe is considered within the weak-field modification of General Relativity involving the cosmological constant. This approach enables to describe the dynamics of groups and clusters of galaxies, to explain the discrepancy in the observational properties of the local (late) and the global (early) Universe, i.e. the Hubble tension as a result of two flows, local and global ones, with non-equal Hubble parameters. The kinetic analysis with the modified gravitational potential involving the cosmological constant is shown to predict semi-periodical structure of filaments in the local universe. In the local scale this complements the Zeldovich pancake theory of evolution of the primordial density perturbations and of structure formation in the cosmological scale. The role of the cosmological constant is outlined in rescaling of the physical constants from one aeon to another within the Conformal Cyclic Cosmology.}

\abstract{The structure formation in the local Universe is considered within the weak-field modification of General Relativity involving the cosmological constant. This approach enables to describe the dynamics of groups and clusters of galaxies, to explain the discrepancy in the observational properties of the local (late) and the global (early) Universe, i.e. the Hubble tension as a result of two flows, local and global ones, with non-equal Hubble parameters. The kinetic analysis with the modified gravitational potential involving the cosmological constant is shown to predict semi-periodical structure of filaments in the local universe. In the local scale this complements the Zeldovich pancake theory of evolution of the primordial density perturbations and of structure formation in the cosmological scale. The role of the cosmological constant is outlined in rescaling of the physical constants from one aeon to another within the Conformal Cyclic Cosmology.}

\section{Introduction}

During years I had the fortunate possibility to discuss with Alexei Starobinsky broad scope of cosmological and physical problems, including those resulted in joint publications dealing with the topology of the Universe \cite{GSt1,GSt2}. Below, the problem of the structure formation will be discussed which was among the topics of research by Yakov Zeldovich \cite{Z}, the renown mentor of Starobinsky. Zeldovich, along with the developing the pancake theory \cite{Z,Arn} of the large-scale structure formation, outlined the efficiency of non-relativistic treatment of the structure formation in the local Universe \cite{Z1}.  

The difference in the descriptions of the early and late Universe can be indicated by the Hubble tension, i.e. the evidence on the discrepancy of the value of the Hubble constant obtained at observations of galactic surveys \cite{R1,R2,R3,Dai1,Dai2,Dai3,Dai4} and of Cosmic Microwave Background  \cite{CMB}.

The non-relativistic description of the local Universe that we consider below is based on a theorem \cite{G} on the general function satisfying the condition of the equivalency of the gravity of the sphere and of a point mass located in its center. That function for the force has the form \cite{G} 
\begin{equation}\label{F}
	\mathbf{F}(r) = \left(-\frac{\alpha}{r^2} + \Lambda r\right)\hat{\mathbf{r}}.
\end{equation}   
The second term in the right-hand-side corresponds to the cosmological term in McCrea-Milne non-relativistic cosmological model \cite{MM}, and the cosmological constant $\Lambda$ enters the weak-field General Relativity. Then, within this $\Lambda$-gravity one has 2 gravity constants, $G$ and $\Lambda$,
and weak-field modified General Relativity with the metric
\begin {equation} \label {mod}
g_{00} = 1 - \frac{2 G m}{r} - \frac{\Lambda r^2}{3}; 
\,\,\, g_{rr} = \left(1 - \frac{2 G m}{r} - \frac {\Lambda r^2}{3}\right)^{-1}.
\end {equation} 

This metric was known before, the Schwarzschild – de Sitter metric \cite{Rind}, as describing a black hole type solution. While the derivation based on the theorem on sphere-point equivalency enables to describe large-scale configurations such as galaxy groups and clusters and their flows.  
The role of $\Lambda$ in Eqs.(1),(2) within group-theoretical approach can be represented depending on the sign of $\Lambda$ as three solutions for Einstein equations corresponding to the isometry groups, as shown in the Table 1 \cite{GS1}.

\begin{table}
	\caption{Isometry groups for different values of $\Lambda$.}\label{tab1}
	\centering
	\scalebox{1}{
		\begin{tabular}{|p{2.9cm}||p{3.4cm}|p{2.4cm}||}
			\hline
			
			Sign& Spacetime&Isometry group \\ \hline
			\hline
			$\Lambda > 0$ &de Sitter (dS) &O(1,4)\\ \hline
			$ \Lambda = 0$ & Minkowski (M) & IO(1,3)\\ \hline
			$\Lambda <0 $ &Anti de Sitter (AdS) &O(2,3)\\ 
			\hline
		\end{tabular}
	}
\end{table}
The crucial point regarding Eq.(1) is that, it satisfies the first condition of Newton's shell theorem, namely the sphere-point equivalence but not the second one, i.e. force-free field inside a spherical shell. One can recall a classical example, when Einstein, to develop the cosmological model, posed two conditions: (1) existence of non-zero mean density of space, (2) the size (radius) of the space does not depend on time \cite{E}. When Friedmann removed the second condition, he got new remarkable solutions which later appeared to be in agreement with the observations.    

  Below, we deal with the non-relativistic description  of the dynamics of the groups and clusters based on Eq.(1), involving the difference of the local flow of galaxies and of the global expansion of the Universe, in the context of the Hubble tension. The local structure formation is described within this non-relativistic approach with a cosmological constant in Eq.(1). Certain consequences of the consideration of $\Lambda$ as a physical constant are also discussed, including in the context of Cyclic Conformal Cosmology (CCC) of Penrose \cite{P}.

\section{Galaxy groups and clusters}

Turning to the application of $\Lambda$-gravity to the galaxy systems, first let us turn to the virilalized systems. For them one can derive from Eq.(1)
the following formula \cite{G1}
\begin{equation}   
	\Lambda=\frac{3\sigma^2}{2 c^2 R^2}\simeq 3\,\, 10^{-52} (\frac{\sigma}{50\, km s^{-1}})^2(\frac{R}{300 \,kpc})^{-2} m^{-2},
\end{equation} 
where $\sigma$ is the velocity dispersion and $R$ the characteristic radius of the system. Using this formula one can test the role of the $\Lambda$-term in Eq.(1) observational data on galaxy systems.  Table 2 \cite{G1}, based on the data in \cite{Kar} of a sample of 17 galaxy groups of Hercules–--Bootes region, with $\it rms$ galactic velocities $\sigma$ and the harmonic average radii $R_h$ of the groups, exhibits the values of $\Lambda$ obtained using Eq.(3); the galaxy groups are denoted by the name of their brightest galaxy. 

\begin{center}
	{\bf Table 2.} $\Lambda$ obtained for the galaxy groups of the Hercules-Bootes region.
	
	{
		\renewcommand{\baselinestretch}{1.2}
		\renewcommand{\tabcolsep}{3.5mm}
		\small
		\begin{tabular}{ | l | r | r | r | }
			\hline
			Galaxy group    &  $\sigma (km/s^{-1})$ & $R_h(kpc)$ & $\Lambda(m^{-2})$\\ \hline		
			\hline
			NGC4736  &   50 & 338 &  3.84E-52 \\ \hline
			NGC4866  & 	 58 & 168 &  2.09E-51 \\ \hline
			NGC5005  & 	114 & 224 &  4.55E-51 \\ \hline
			NGC5117  & 	 27 & 424 &  7.12E-53 \\ \hline
			NGC5353  & 	195 & 455 &  3.23E-51 \\ \hline
			NGC5375  & 	 47 &  66 &  8.91E-51 \\ \hline
			NGC5582  & 	106 &  93 &  2.28E-50 \\ \hline
			NGC5600  & 	 81 & 275 &  1.52E-51 \\ \hline
			UGC9389  & 	 45 & 204 &  8.55E-52 \\ \hline
			PGC55227 & 	 14 &  17 &  1.19E-50 \\ \hline
			NGC5961  & 	 63 &  86 &  9.43E-51 \\ \hline
			NGC5962  & 	 97 &  60 &  4.59E-50 \\ \hline
			NGC5970  & 	 92 & 141 &  7.48E-51 \\ \hline
			UGC10043 & 	 67 &  65 &  1.87E-50 \\ \hline
			NGC6181  & 	 53 & 196 &  1.28E-51 \\ \hline
			UGC10445 & 	 23 & 230 &  1.76E-52 \\ \hline
			NGC6574  & 	 15 &  70 &  8.07E-52 \\ \hline
			\hline
			Average  &      &     &  8.24E-51 \\ \hline
			St.deviation	   &      &     &  1.15E-50 \\ \hline
			\hline
		\end{tabular}
	}
\end{center}
The value of the cosmological constant obtained from CMB Planck data yields $\Lambda = (1.09 \pm 0.028) \times 10^{-52} m^{-2}$ \cite{CMB}.
The correspondence of the latter value of $\Lambda$ with those in Table 2 is visible. In \cite{GS1,GS2} various samples of galaxy groups and clusters, including groups of Leo/Cancer region, clusters of CLASH, LoCuSS Surveys, Planck clusters,  are considered and it is shown that $\Lambda$-gravity with Eq.(1) does agree with those observational data. 

\section{Local and global Hubble flows: Hubble tension}

The $\Lambda$-gravity described above enables to address the Hubble tension \cite{R1,R2,R3,Dai1,Dai2,Dai3}. Consider the following two equations which although look similar but have rather different content \cite{GS3a,GS3,GS4} 
\begin{equation}\label{Hl}
	H_{local}^2 = \frac{8 \pi G \rho_{local}}{3} + \frac{\Lambda c^2}{3},
\end{equation}
\begin{equation}\label{Hg}
	H_{global}^2 = \frac{8 \pi G \rho_{global}}{3} + \frac{\Lambda c^2}{3}.
\end{equation}
The first equation is derived using Eq.(1) within non-relativistic description of the local Universe, while the second one is the Friedmannian equation for FLRW metric. The discrepancy in the values of the relevant Hubble parameters, $H_{local}$ and $H_{global}$, due to the different values of the mean densities, is then evident, thus suggesting a natural solution to the Hubble tension.

Then, Eq.(1) enables to define a \textit{critical radius} where the $\Lambda$-term in the r.h.s. dominates upon the first term and the test particle will run away the central body
\begin{equation}\label{rcrit}
	r_{crit}^3 = \frac{3 GM}{\Lambda c^2}. 
\end{equation}
This scale attributed to the local Hubble flow, in \cite{GS3,GS4} is shown to agree with the data on the galactic flows in the vicinity of the Local Group, Virgo cluster, Virgo Supercluster and Laniakea Supercluster.

One can also obtain {\it absolute} constraints on the lower and upper values for the local Hubble parameter \cite{GS4},
\begin{equation}
H_{min} = \sqrt (\Lambda c^2/3) = 56.2\, km/sec\,\, Mpc^{-1},
\end{equation}
\begin{equation}
H_{max} = \sqrt (\Lambda c^2) = 97.3\, km/sec\,\, Mpc^{-1}. 
\end{equation}
The lower constraint is obtained from the condition of the validity of the weak-field limit, while the upper limit is the value of Hubble parameter at $r_{crit}$ (for details see \cite{GS4}). These constraints are in full agreement with the observational data on the galactic flows in the local Universe.

\section{Structure formation}

As mentioned above, Zeldovich pancake theory \cite{Z,Arn} considers the evolution of primordial fluctuations to produce the large-scale
cosmological structures. On the local scale,  the self-consistent gravitational interaction can be crucial to ensure the formation of structures on that scale.
Hence, in \cite{GFC1,GFC2,GFC3} the non-relativistic (non-GR) kinetic Vlasov mechanism of the formation of cosmic voids and walls in considered. Crucial point is the
use of Eq.(1) for the gravitational interaction of the particles, i.e. the existence of repulsive $\Lambda$-term along with the gravitational attraction. 

Then the Vlasov--Poisson equations to describe the system of $N$ particles is given as (see \cite{GFC1,GFC2,GFC3} for details)
\begin{equation}
\frac {\partial F({\bf x},{\bf v},t)}{\partial t} +
{\rm {div}}_{\bf x}({\bf v}F)+\widehat {G}(F;F)=0,
\end{equation}
\begin{equation}
\widehat{G}(F; F) \equiv	-\big ({\rm{\nabla}}_{\bf v}F\big )
	\big({\nabla }_{\bf x}(\Phi [F({\bf x})]\big),
	\label{1}
\end{equation}
\begin{equation}
	{\Delta}_{\bf x}^{(3)}\Phi[F({\bf x})]\big|_{t=t_0}=
	A G \int F( {\bf x},{\bf v},t_0)\:d{\bf v} - \frac{c^2\Lambda}{2},
	\label{2}
\end{equation}
where $F({\bf x},{\bf v},t)$ is the distribution function of particles interacting
according to Eq.(1), $A$ is a normalization factor for particle density.

The analysis of this system of equations leads to semi-periodic solutions which for the normalized radial coordinate $u({\bf x})$ have the form \cite{GFC2}
\begin{equation}
	u({\bf x}) =\sum_{m=0} \big(\eta\exp(-U^{(0)}) \big)^{m+1/2} \bigg( \sum_{ \ell =1}^{m} {}_{(I)}C^{(2m+1)}_{2\ell + 1} 
\end{equation}
$$
\sin\big((2\ell+1){\bf q }{\bf x}\big) +
{}_{(III)}C_{2m+1} \sin\big( {\bf q}{\bf x} \big)
\big) +
$$
$$
+\sum_{m=1}^\infty \big(\eta\exp(-U^{(0)}) \big)^m \bigg( \sum_{\ell =1}^{m} {} _{(II)}C^{(2m)}_{2\ell}\cos\big(2m {\bf q}{\bf x} \big) +
$$
$$
{}_{(III)}C_{2m} \sin\big( {\bf q}{\bf x} \big) \bigg).
$$
As shown in \cite{GFC1,GFC2,GFC3} such semi-periodic solutions of the Vlasov-Poisson equations appear due to the repulsive $\Lambda$-term in the interacting potential, thus providing a mechanism for formation of voids, walls, and 1D and 2D-filaments. The critical radius given by Eq.(6), and hence, the interplay of $\Lambda$ and of the local matter density determines the scale of the voids and appears to be in agreement with the observational data in the local Universe \cite{GFC2,GFC3}.
 
\section{$\Lambda$ as a fundamental constant}  

The $\Lambda$-gravity considered above based on the sphere-point equivalence theorem \cite{G}, suggests two gravity constants, $G$ and $\Lambda$ in Eq.(1).
If so, then one can go further and consider the cosmological constant $\Lambda$ among fundamental physical constants, along with $G,c, \hbar$. Note, that Einstein has denoted as {\it universal constant} the cosmological constant when he introduced it in his cosmological model \cite{E,E1}. 

Then, the set of (3+1) constants ($G, \Lambda, c, \hbar$) enables one to have the following dimensionless quantity \cite{GS5}.
\begin{equation}\label{dimless}
	I=\frac{c^{3a}}{\Lambda^a G^a \hbar^a},
\end{equation}
where $a$ is a real number, while no dimensionless ratio is possible to construct from the classical 3-set ($G, c, \hbar$). 

The relation Eq.(13) of the 4 constants within a numerical factor and $a=1$ coincides with that of the entropy of de Sitter event horizon \cite{Bek,HG,P}
\begin{equation}\label{dSE}
	I_{dS}= 3 \pi \frac {c^3}{\Lambda G \hbar},
\end{equation}
and with the Bekenstein bound \cite{BekB} for the information in de Sitter space 
\begin{equation}\label{BBdS}
	I_{BB} =  \frac {3 \pi c^3}{\Lambda G \hbar ln 2}.
\end{equation}

The consideration of $\Lambda$ as a physical constant has consequences for the  Conformal Cyclic Cosmology (CCC) \cite{P}, for which the second law of thermodynamics and the positive $\Lambda$ are basic conditions. Then, the dimensionless ratio Eq.(13) becomes important condition regarding the information transfer from one aeon to the next one (see \cite{P,GP1,GP2}), as invariant of conformal transformation
\begin{equation}
	\tilde{g}_{\mu\nu}=\Omega^2 g_{\mu\nu}.
\end{equation}

Namely, as shown in \cite{GS5}, then the physical constants can be rescaled from one aeon to the successive one
\begin{equation}
	c\to a_1 c, \quad \hbar \to a_2 \hbar, \quad G\to a_3G, \quad \Lambda \to a_4 \Lambda, \quad a_i \in R^{+},
\end{equation}
keeping the condition
\begin{equation}
	\frac{a_1 ^3}{a_2 a_3 a_4}=1,
\end{equation}

Thus, considering $\Lambda$ among 3 other physical constants will imply within CCC: (a) the same global cosmological dynamics in the aeons, (b) rescaling of the physical constants with modifications for physical processes from one aeon to another.

\section{Conclusions}

The observational indications on the peculiarities in the properties of the early and late Universe stimulate consideration of non-uniform mechanisms responsible for the relevant structure formation. The Zeldovich pancake theory \cite{Z} of structure formation at global scale deals with the evolution of the primordial density perturbations, while at local scales the self-consistent gravitational interaction can have dominating contribution. 

We concisely discussed the properties of the local Universe and the formation of the matter structures based on the weak-field modification of General Relativity with the cosmological constant $\Lambda$. That approach follows from the theorem \cite{G} on sphere-point gravity identity when the relevant general function includes a $\Lambda$-term. The Hubble tension then is naturally explained as a result of local and global flows with non-identical Hubble parameters. The analysis of Vlasov-Poisson equation set with the potential involving the $\Lambda$-term leads to the prediction of voids and walls in the matter structure of the local Universe, so that the cosmological constant, along with the local matter density, defines the scale of the voids.

Forthcoming more deep observational surveys in acquiring accurate comparative features of the early and late Universe can enable one to reveal the relevant physical processes in more details.        

\begin{acknowledgement}
I am thankful to A.Stepanian for the efficient joint work. 
\end{acknowledgement}


\end{document}